\documentclass[preprint,aps,prl,epsf]{revtex4}   
\usepackage{amsmath}
\usepackage{amssymb}

\usepackage{epsf,graphicx}
\include{epsfx}
\usepackage{xtab}
\usepackage{color,soul}
\usepackage{soul}
\definecolor{HPgray}{gray}{0.50}
\definecolor{CYSgray}{gray}{0.20}

\usepackage[labelfont=bf,labelsep=period,justification=raggedright]{caption}

\begin{document}

\title{\textbf{Molecular jamming -- the cystine slipknot mechanical clamp in all-atom simulations}}
\author{{\bf {\L}ukasz Pep{\l}owski$^1$, Mateusz Sikora$^2$, Wies{\l}aw Nowak$^1$, and Marek Cieplak$^2$}}

\address{
$^1$ Institute of Physics, Nicolaus Copernicus University, Torun, Poland\\
$^2$ Institute of Physics, Polish Academy of Sciences,
Warsaw, Poland\\}

\begin{flushleft}
$\ast$ E-mail:  mc@ifpan.edu.pl
\end{flushleft}

\vskip 40 pt

\begin{abstract}
A recent survey of 17 134  proteins has identified a new class of proteins
which are expected to yield stretching induced force-peaks in the range
of 1 nN. Such high force peaks should be due to forcing of a slip-loop
through a cystine ring, i.e. by generating a cystine slipknot. The survey
has been performed in a simple coarse grained model. Here, we
perform all-atom steered molecular dynamics simulations on 15 cystine knot
proteins and determine their resistance to stretching.
In agreement with previous
studies within a coarse grained structure based model, the level of resistance
is found to be substantially higher than in proteins in which the mechanical
clamp operates through shear. The large stretching forces arise through
formation of the cystine slipknot mechanical clamp and the resulting
steric jamming. We elucidate the workings
of such a clamp in an atomic detail. 
We also study the behavior
of five top strength proteins with the shear-based mechanostability
in which no jamming is involved.
We show that in the atomic model, the jamming state is relieved
by moving one amino acid at a time and there is a choice in the selection
of the amino acid that advances the first. In contrast, the coarse grained model
also allows for a simultaneous passage of two amino acids.
\end{abstract}

\maketitle

\section{Introduction}
Single-molecule manipulation \cite{COSTB,encyclopedia,Crampton} has opened new
perspectives on understanding of the mechanical processes taking place
in a biological cell and may offer insights into design of nanostructures
and nanomachines.
Examples of manipulation of a protein include stretching \cite{C3},
mechanically controlled refolding \cite{FernandezLi}, knot
untying \cite{Alam} and knot tightening \cite{Rief}. The experimental
studies pertained to only a handful of systems and yet demonstrated
richness of possible mechanical behaviors. Experiments on stretching generate
information on mechanostability. It can be captured by providing the value
of $F_{max}$ -- the largest force that is needed to unravel the tertiary
structure of a protein. $F_{max}$ can be as large as 480 pN  -- the value
measured for scaffoldin c7A \cite{Valbuena}. The structural motif that is the core
of resistance to stretching is known as the mechanical clamp. In most
cases of large mechanostability, the mechanical clamp consists primarily of
parallel \cite{Lu1999,B2} (and sometimes antiparallel \cite{JPCM}) $\beta$-strands
that get sheared on pulling at two points of attachment, which are often
the terminal amino acids. This mechanism is operational, for instance,
in scaffoldin \cite{Valbuena}  and titin \cite{Lu1999,B2}.

\begin{figure}[!ht]
\epsfxsize=3.2in
\centerline{\epsffile{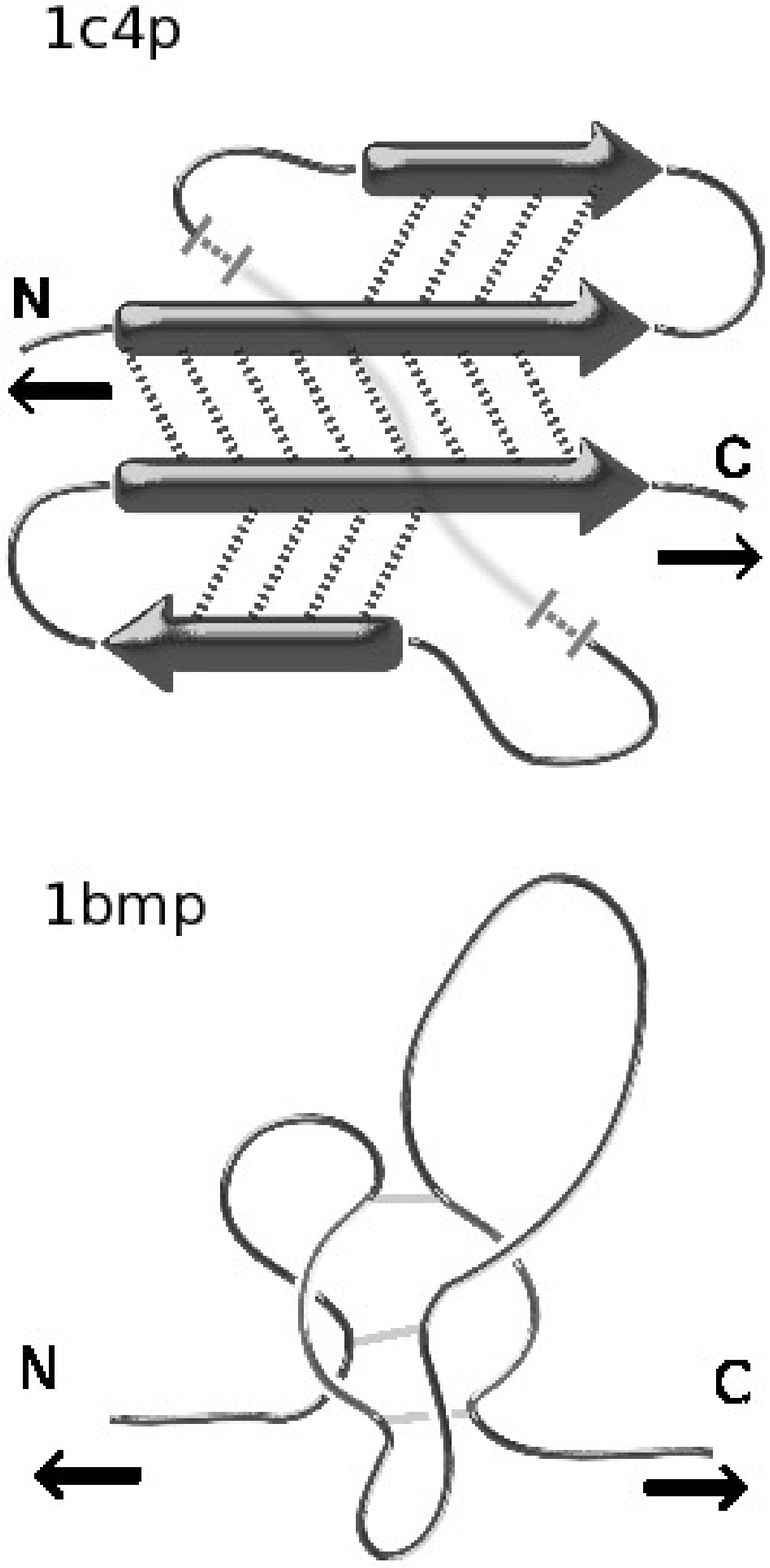}}
\vspace*{-1cm}
\caption{{\bf Two basic classes of the mechanical clamps.}
The top panel shows the shearing mechanical clamp as illustrated
for protein 1c4p. The bottom panel illustrates the cystine slipknot
mechanical clamp in the case of protein 1bmp.
}
\label{clamps}
\end{figure}

Recently, we have made stretching simulations for 17 134 proteins within a coarse
grained structure-based model \cite{plos2009}. This survey pertained to
proteins built of no more than 250 amino acids.
Its results are deposited at the BSDB database
(Biomolecule Stretching Database, www://info.ifpan.edu.pl/BSDB/)  as described
in ref. \cite{BSDB}. The survey not only identified
about 35 proteins of larger mechanostability than the scaffoldin but it
also discovered an entirely new type of the mechanical clamp: the cystine
slipknot (CSK). In fact, the 13 top strongest model proteins are endowed
with this mechanism.
It involves forced dragging of a piece of the backbone,
known as the slip-loop, through a cystine ring which is a
special case of a knot-loop. Thus CSK does not
exist in the native state but it develops through pulling.
The cystine ring arises by connecting two backbone segments by two
disulfide bonds between two pairs of cysteins. The ring is a part of what is
known as the cystine knot  (CK) motif \cite{Craik,CKstability,Craik2} and has
been first discovered in the superfamily of growth
factors \cite{odkrywcy,pallaghy,Craik1} and is
responsible for the high mechanical stability of collagen \cite{Moroder}.
It typically contains between 8 and 15 residues \cite{Craik} and it has
emerged probably as a result of convergent evolution \cite{convergent}.
The pulling-induced
dragging takes place because of the presence of a third disulfide
bond that pierces the ring (Figure \ref{clamps}).
This CSK mechanism has been predicted \cite{plos2009} to yield $F_{max}$
exceeding even 1000 pN. Such large forces would arise because of
jamming resulting from the slip-loop getting stuck in the
cystine ring. An identifiable force peak maximum appears if the steric
hindrances are overcome and the clinch is released.
It should be noted that the acronym CSK has been initially
meant to stand for "cysteine slipknot" \cite{plos2009} as the motif involves cysteins.
However the term "cystine slipknot" used in this paper seems more apt
as it brings the role of the disulfide bonds into fore.

At present, there are three well established families within
the cystine knot superfamily: growth factor cystine knots
(GFCK, e.g. vascular endothelial growth factor \cite{Iyer,Well}),
inhibitor cystine knots (ICK, e.g. scorpion venom proteins \cite{Chiche}) and
cyclic cystine knots (CCK) which have no terminal amino acids \cite{Gruber}.
The cystine knot proteins studied in this work are members of the first family.
The CK proteins are mostly extracellular factors and they are
known for their remarkable stability
against enzymatic cleavage \cite{CKstability,Craik2}.

These 35 strongest proteins, with the CSK mechanical clamp and without,
have not yet been studied in single molecule stretching experiments.
Furthermore, our predictions are based on a
simple model that accounts for the geometry and size of the side
groups in a very approximate manner (primarily through the length
parameters in the contact potentials) whereas the success in dragging through the cystine
ring crucially depends on the actual size and structure of the moving
segments involved. In this paper, we perform all-atom steered molecular
dynamics (SMD) simulations and confirm the existence of the CSK mechanical
clamps with the values of $F_{max}$ being substantially larger than
those associated with the shear-based mechanical clamps. We focus
on the top 20 proteins identified in the survey as listed in Table 1.
We also consider titin which serves as a benchmark
\cite{Eric,Linke,T1,Kell,R4,C6,Fernandez,Mar1,F1}.

We find that even though the coarse grained model captures the essence
of the CSK clamp and, in particular, that it involves jamming,
it misses several interesting atomic-level components.
The first component is the different origin of two possible stretching pathways:
in the all-atom model what matters is the direction of approach of the
pulled backbone to the cystine ring whereas it is the number of amino acids
(one or two) that cross the ring simultaneously in the coarse grained model.
Another component is that the tendency of developing multiple force
peaks on a given pathway is stronger in the all-atom model because
size modulations in the side groups may split the force peaks.
These modulation are also relevant for thermal stability of
the CSK-containing proteins (CSK proteins, for short).

The SMD simulations need to fit in the time scale of several tens of
ns and thus have to be performed at pulling speeds which are
orders of magnitude greater than what can be achieved in experiments
or in coarse grained simulations.
The basic pulling velocity used here is  0.04 {\AA}/ps.
This necessarily leads to the
values of $F_{max}$ which are much larger than experimental, typically
an order of magnitude larger \cite{Lu1999}.
Thus our results on $F_{max}$ are also overblown for this reason and have
to be scaled down. We show that the functional form of this scaling down
is complicated as it involves a crossover between a logarithmic and linear
speed dependence. The latter is expected at high speeds when the drag force
dominates. Nevertheless, the values of $F_{max}$ obtained suggest
that CSK comes with a much larger mechanostability than the shear-based
clamp also at the experimental speeds.

For many CSK proteins there is a considerable variation
between trajectories in the values of $F_{max}$.
Pinpointing a specific force-based ranking of the proteins is
difficult without generating significant statistics of the
trajectories. It need not coincide with the one obtained by using
the coarse grained model (and based on the dominant pathways).
On the other hand, the ranking of the proteins with the shearing
mechanical clamp, and listed in Table 1, is the same.
In addition to providing the force-displacement ($F-d$) trajectories
for typically two trajectories for each protein studied, we also
characterize the geometry and sequential make-up of the resulting
mechanical clamps and study
rupturing of the hydrogen bonds.

\section{Results and Discussion}
\subsection{The force-displacement curves}
Table 1 lists 21 proteins that are studied in this paper.
The proteins there are arranged according to the value of $F_{max}$
as obtained in the coarse grained model \cite{plos2009}. The energies
and forces of that model are determined in terms of an energy parameter
$\epsilon$ which measures the depth of the potential well in the
Lennard-Jones-like native contacts. The forces written in Table 1 are
converted to the experimental units using  $\epsilon$/{\AA}=110$\pm$30 pN --
the result obtained by comparing the theoretically determined values of
$F_{max}$ to the established experimental data \cite{plos2009}.
Notice a considerable error bar in the conversion factor. Another reason
for not treating the protein ranking in Table 1 in a verbatim
way is that for the CSK proteins
at the high end of the force, there is a considerable
trajectory dependence in the values of $F_{max}$  both in the
coarse grained and all-atom models. Even though the coarse
grained values in Table 1 are  obtained through averaging
over 10 trajectories, a substantial dispersion in the CSK cases remains.
The important point is that
among the 21 proteins listed in Table 1, 15 have the CSK mechanical
clamp as determined in the coarse grained model. They are ranked 1-13 and then
16 and 18. There are more than 100 other CSK proteins in the full set of 17 134
considered in ref. \cite{plos2009} but they are expected to correspond to
lower forces, at least within the coarse grained model.

\begin{figure}[!ht]
\epsfxsize=6in
\centerline{\epsffile{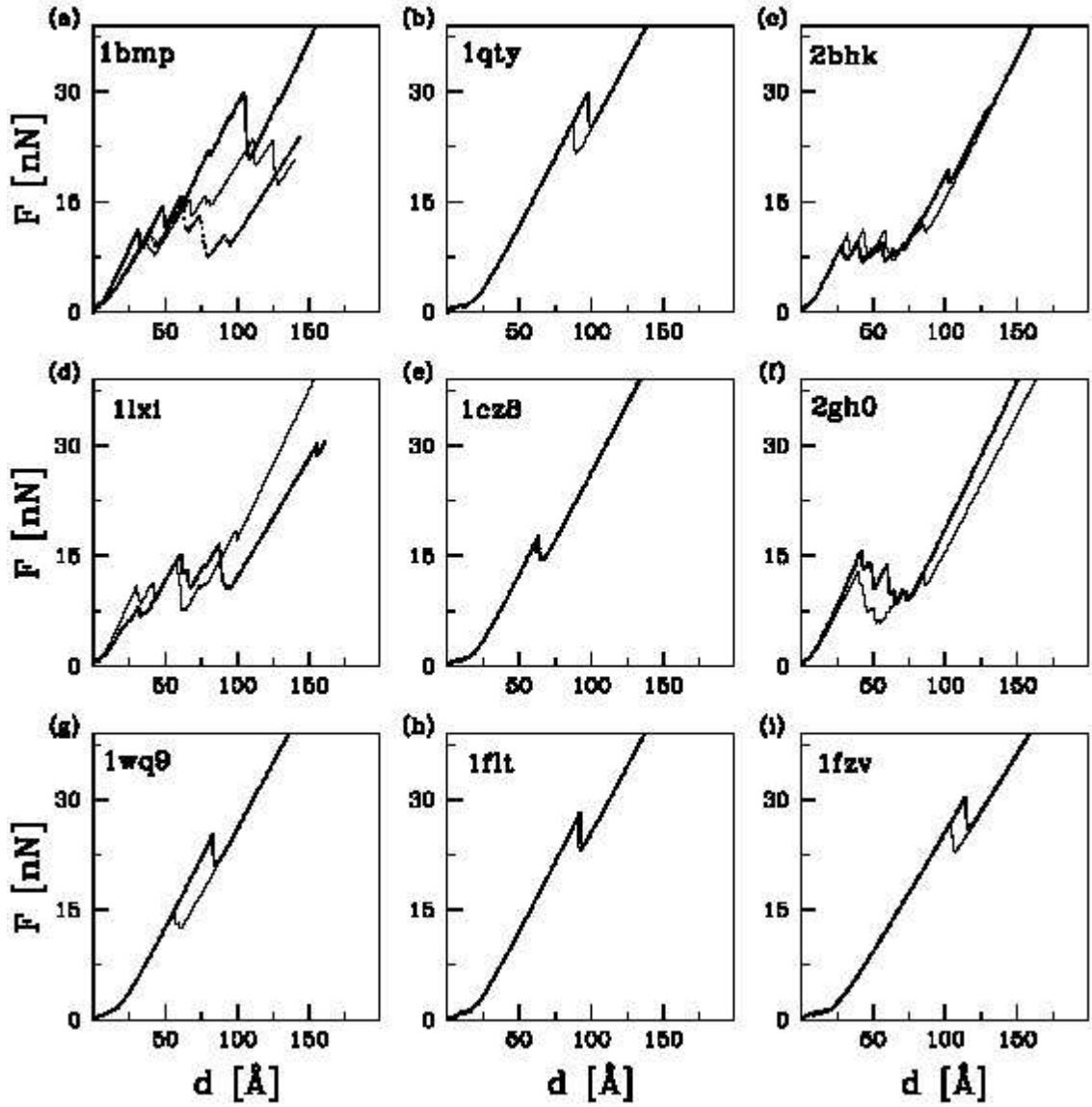}}
\caption{{\bf Examples of the all-atom $F-d$ trajectories for the top 9 proteins
listed in Table 1.}
}
\label{three}
\end{figure}

\begin{figure}[!ht]
\epsfxsize=6in
\centerline{\epsffile{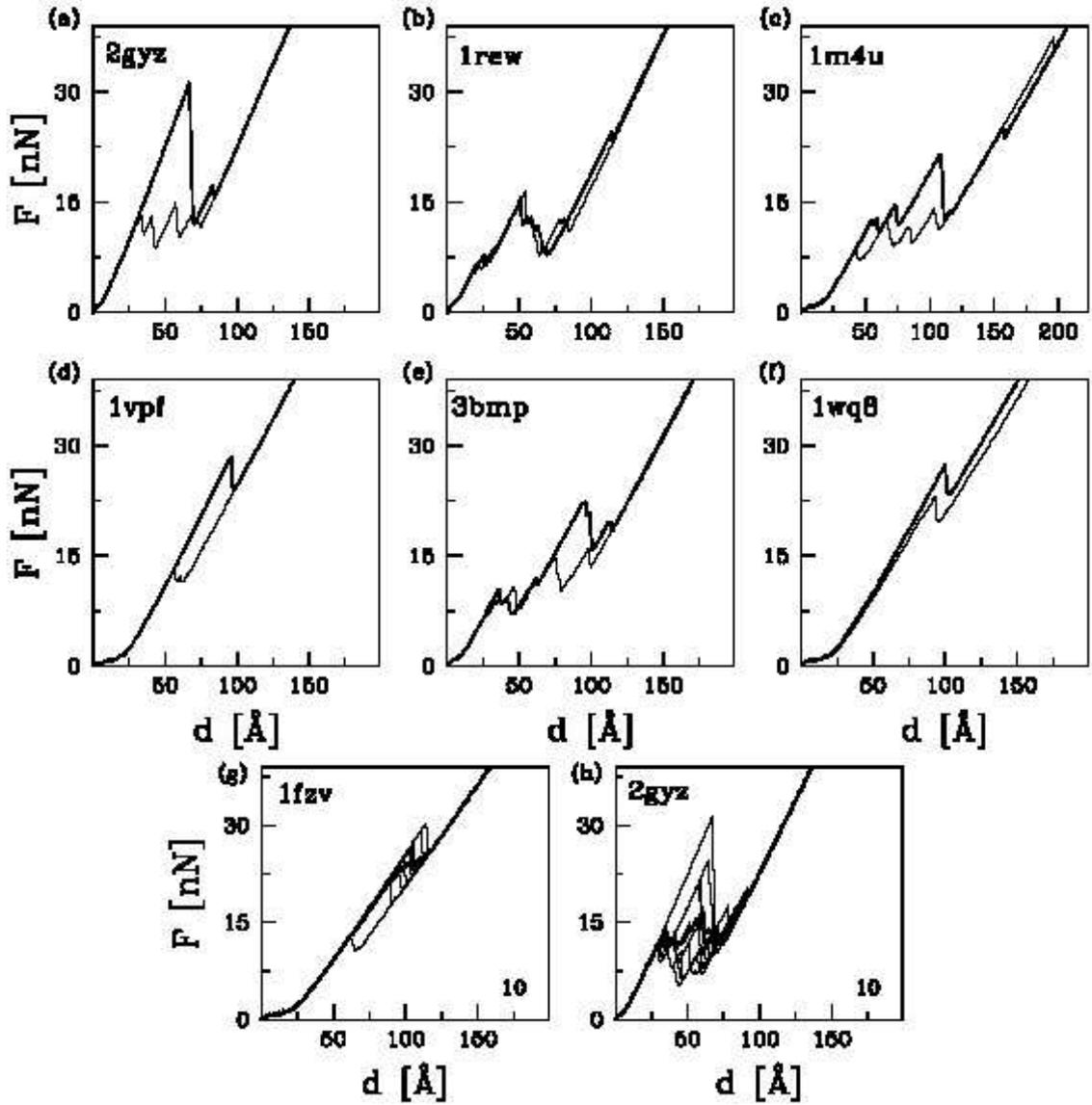}}
\vspace*{1cm}
\caption{{\bf Examples of the $F-d$ trajectories for the CSK proteins.}
The top 6 panels show examples of two all-atom $F-d$ trajectories
for the proteins ranked
as number 10, 11, 12, 13, 16, and 18 in Table 1. Panels (g) and (h)
(with 10 written in the lower right corner) show ten
$F-d$ trajectories for  1fzv and 2gyz as indicated. The thick solid line
corresponds to the average over the ten trajectories.
}
\label{next}
\end{figure}

\begin{figure}[!ht]
\epsfxsize=6in
\centerline{\epsffile{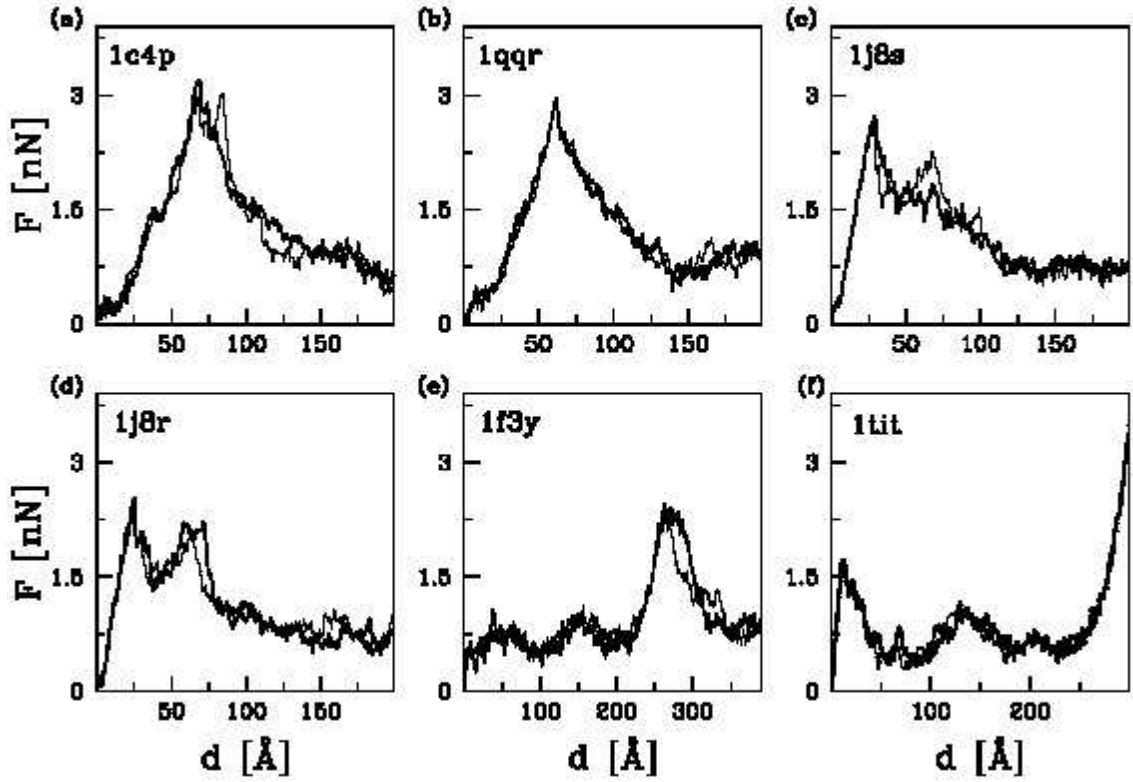}}
\vspace*{1cm}
\caption{{\bf Examples of two $F-d$ trajectories for
proteins are endowed with the shear mechanical clamps.}
These proteins are the six remaining entries
that are listed in Table 1.
}
\label{shear}
\end{figure}

The protocol of our all atom simulations is explained in the Methods section.
The resulting $F-d$ plots are displayed in Figs.\ref{three} through
\ref{shear}.
Full extension is never reached as this would require breaking
disulfide bonds.
Figures \ref{three} and \ref{next} refer to the CSK proteins
whereas Fig.\ref{shear} to the proteins with the shear mechanical clamp.
In the latter case, the values of $F_{max}$ are an order of magnitude smaller
and the trajectory dependence is substantially weaker. Previous
all-atom studies \cite{Lu1999} for titin indicate that $F_{max}$ obtained
at such speeds is an order of magnitude higher than the experimental
value of 204 pN \cite{R4,C6}.

Among the 15 CSK proteins, five (1cz8, 2gh0, 1rew, 1m4u, and 3bmp) have both
trajectories corresponding to $F_{max}$ of order 15 nN whereas the remaining
ten have at least one in the range of 30 nN  as in the case of
1bmp (bone morphogenic protein) which is the mechanostability
leader in the coarse grained evaluation.
Larger statistics of ten trajectories have been obtained for two proteins,
1fzv and 2gyz, as shown in panels (g) and (h) of Fig.\ref{next}.
They indicate that $F_{max}$ takes values in the whole range in which
the upper reach is about twice as large as the lower reach.
2gyz is seen to have more trajectories with forces lower than 1fzv
whereas in the coarse grained model the two systems appeared to
behave similarly.

All of these force values are significantly larger than those associated with
the shearing clamp shown in Fig.\ref{shear}. The largest value of
$F_{max}$ in this group is $\sim$ 3 nN as observed for
the homologous pair 1c4p and 1qqr (panels (a) and (b) in Fig.\ref{shear}).
Despite the spread in $F_{max}$ between the trajectories
for the CSK proteins, many $F-d$ plots look essentially similar indicating existence
of a single pathway. However, some proteins, like  2gyz
(panels (a) and (h) of Fig.\ref{next}) come with two distinctive
patterns and thus two pathways: one with two force peaks and
$F_{max}$ exceeding 30 nN and another with four peaks and $F_{max}$
of order 15 nN. We shall discuss why these differing patterns arise later on.

\begin{figure}[!ht]
\epsfxsize=6in
\centerline{\epsffile{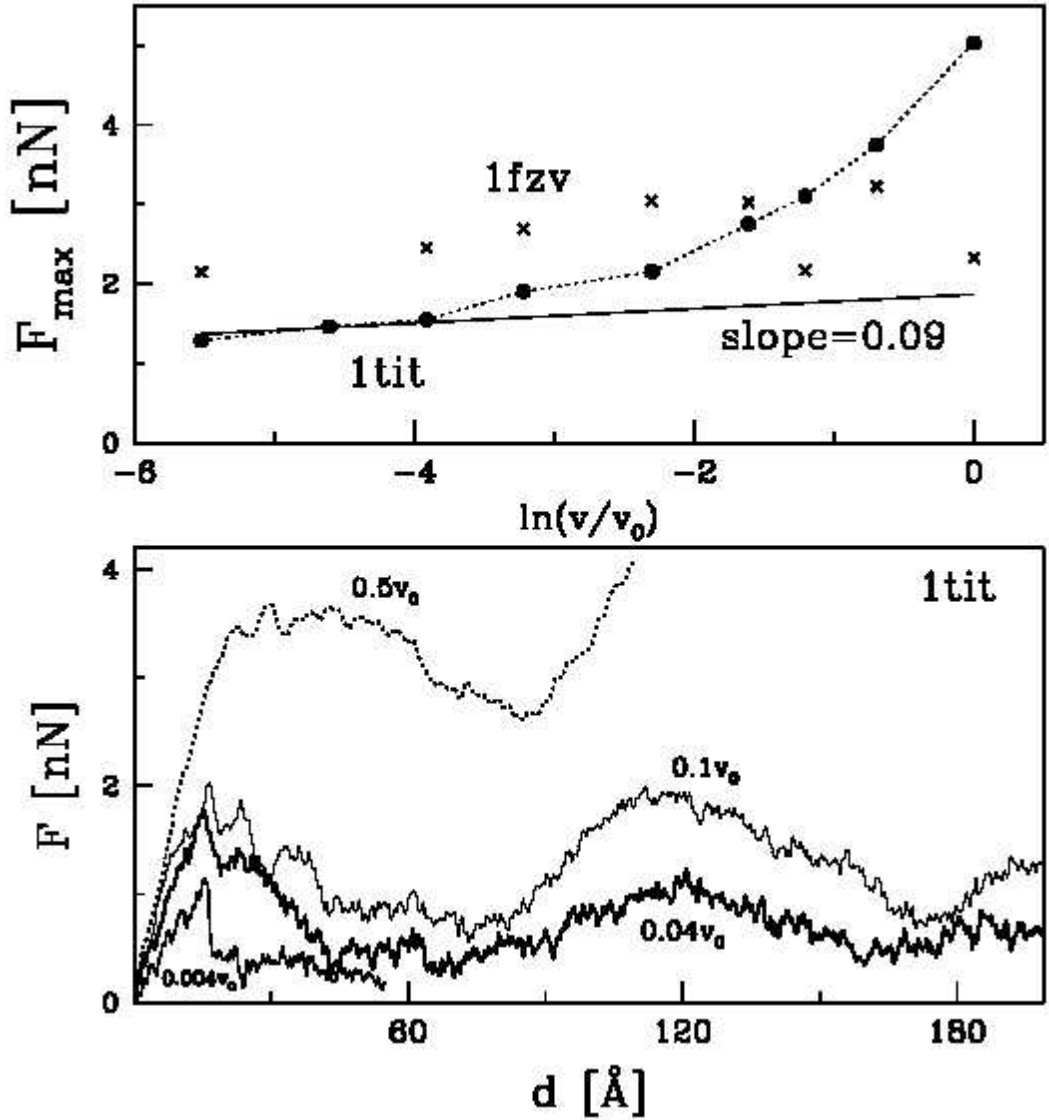}}
\vspace*{1cm}
\caption{{\bf The velocity dependence of $F_{max}$.} The velocity $v_0$ corresponds to
1 {\AA}/ps. The bottom panel shows the $F-d$ plots for titin at the indicated pulling speeds.
The solid data points in the top panel correspond to
$F_{max}$ in the first force peak for titin as a function of
$ln(v/v_0)$, where $v$ denotes the pulling speed.
The data points correspond to single trajectories.
The straight line illustrates a perfect logarithmic dependence.
The slope indicated would yield the experimental value of 204 pN at 600 nm/s.
The crosses correspond to $0.1 F_{max}$ in 1fzv. The three smallest speed
data correspond to single trajectories and the remaining points are averaged
over two trajectories.
}
\label{speed}
\end{figure}

Fig.\ref{speed} addresses the issue of the pulling rate dependence
for titin (1tit) and human placenta growth factor (1fzv). The bottom
panel shows the $F-d$ curves obtained at several speeds: the bigger
the speed the higher the curve and the higher the second force maximum.
We focus on the first maximum as its physics is governed by the
well studied shear mechanical clamp. The corresponding values of
$F_{max}$ are displayed in the top panel. They suggest existence
of a gradual crossover from the logarithmic to linear dependence
on the speed in the regime of the pulling rates studied.
At the slowest speeds, the data points appear to be consistent
with a proper logarithmic extrapolation to forces obtained
experimentally. Nothing is known about the speed dependence of $F_{max}$
for the CSK proteins. The top panel of Fig.\ref{speed} displays
results for 1fzv based on one (for smaller speeds) or two trajectories.
The statistics involved are too poor to guess the functional dependence,
however, the titin-like logarithmic dependence on going to
the experimental speeds remains a possibility. Thus scaling down
of our results for the CSK proteins using the titin-like
logarithmic dependence seems to be a reasonable first guess.
If so then $F_{max}$ of 30 nN may probably correspond to the
experimental 1500 pN. Such forces are comparable to
those needed to rupture covalent N-C and C-C bonds:
1500 and 4500 pN respectively \cite{grandbois}.
On the other hand, the coarse grained simulations suggested values
which would not exceed 1100 pN. At this stage, the coarse grained
model is probably more reliable in this respect.

It should be noted that the coarse grained simulations have been performed
at speeds much closer to those expected at the experimental conditions.
For the CSK proteins, the largest $F_{max}$
obtained is about 5 times larger than the $F_{max}$ for titin.
At the speeds considered in the all-atom simulations, the derived
forces are typically 10 - 20 times larger than for titin.
The extra factor of  2 - 4 compared to the coarse grained model
should be attributed to the different resolution of the two models
and thus to the different ways in which the cystine ring is penetrated.
As will be discussed later, the atomically represented ring
distorts to an oblique form significantly and allows for penetration
to occur only at an acute angle. This kind of penetration is
sensitive to what atoms are facing the plane of the ring.
In contrast, the cystine ring in the coarse grained model remains
more circular and penetration is more vertical.
We do not expect any other changes in the physics of dragging a segment of
the backbone through the cystine ring to be relevant.

\subsection{The mechanical clamp involving shearing of $\beta$-strands}

In order to set the stage for the discussion of the working of the CSK
mechanical clamp,
we first discuss the shear clamp. The largest value
of $F_{max}$  within this mechanism is predicted to arise in the blood clotting
streptokinase $\beta$-domain -- 1c4p (we consider chain A) and its very close
companion hydrolase activator streptokinase domain B -- 1qqr. A sequential
alignment indicates that 1c4p has two more residues at the N terminus, 1qqr
has three more at the C terminus and there are four sites at which the sequences
differ. The high sequence identity results in a very close structural similarity.

\begin{figure}[!ht]
\epsfxsize=3.5in
\centerline{\epsffile{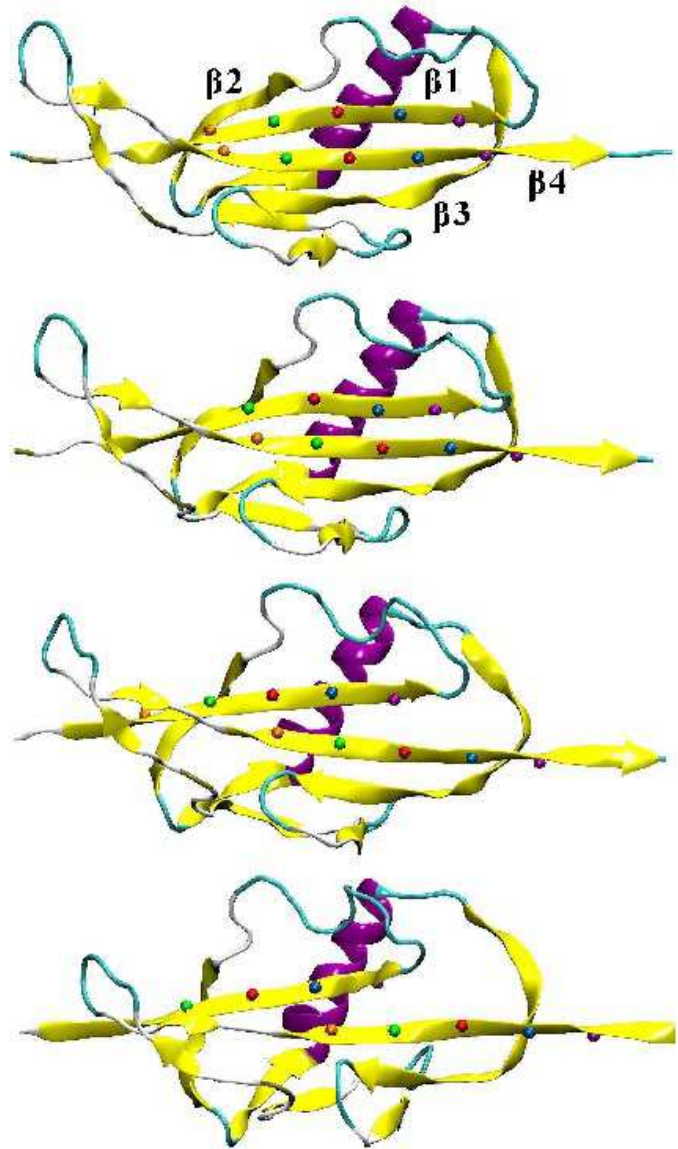}}
\vspace*{1cm}
\caption{{\bf Rupturing of the mechanical clamp in 1c4p through shear.}
The trajectory used is one indicated by the thicker line in Fig. \ref{shear}.
The time frames correspond to 
$d$ of 67, 71, 75, and 83 {\AA}
top to bottom respectively. $F_{max}$ arises at $d$=67 {\AA}.
The $\beta$-strands 1 through 4 correspond to the segments 158--168, 183-188, 214--226,
and 266--278 respectively. The whole structure spans amino acids 149--285.
}
\label{1c4pstory}
\end{figure}

The $F-d$ curves for the two proteins are shown in the first two panels in
Fig.\ref{shear}. The values of $F_{max}$ show only a small spread
between two trajectories. They are 3171 and 3315 pN for 1c4p and 3048
and 3037 for 1qqr indicating a bit larger mechanostability of the former.
In both cases, the mechanical clamp corresponds to the schematic representation
as shown in the top panel of Fig.\ref{clamps}.
The way this mechanical clamp works is shown in  more details in
Fig.\ref{1c4pstory} for 1c4p at four stages of unraveling.
The major shearing action takes place
between the parallel and long $\beta$-strands $\beta_1$ (11 amino acids) and
$\beta_4$ (13 amino acids). There is extra shear between strands $\beta_1$ and
$\beta_2$ as well as between $\beta_3$ and $\beta_4$.

\begin{figure}[!ht]
\epsfxsize=4.6in
\centerline{\epsffile{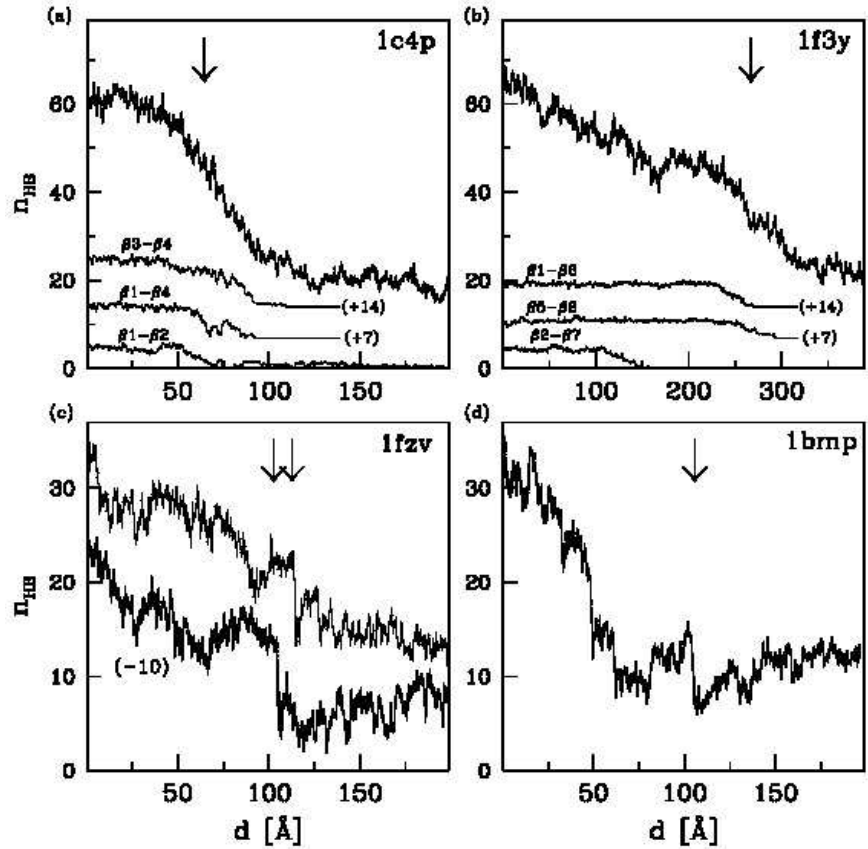}}
\vspace*{1cm}
\caption{{\bf The number of hydrogen bonds as a function of $d$ in typical
trajectories.} The arrows indicate locations of the force peaks.
The upper two panels correspond to proteins with the shearing mechanical
clamps. The top lines show the total number of the hydrogen bonds.
The lower lines represent hydrogen bonds between
the indicated $\beta$-strands. For clarity, they are shifted by the numbers
indicated in the brackets. The sequential segments corresponding to the
$\beta$-strands in 1c4p are listed in the caption of Fig.\ref{1c4pstory}.
For 1f3y, the definitions of the segments are as follows: $\beta 1$ 16--22,
$\beta 2$ 27--33, $\beta 5$ 70--75, $\beta 6$ 105--112, $\beta 7$ 131--137.
The lower panels correspond to the proteins with the
cystine slipknot mechanical clamps. The thicker lines show the total number of
the hydrogen bonds corresponding to the trajectories indicated by the
thicker lines in Fig.\ref{three}. The thinner line for 1fzv corresponds to
the second trajectory generated for this protein.
}
\label{wodor}
\end{figure}

\begin{figure}[!ht]
\epsfxsize=4.5in
\centerline{\epsffile{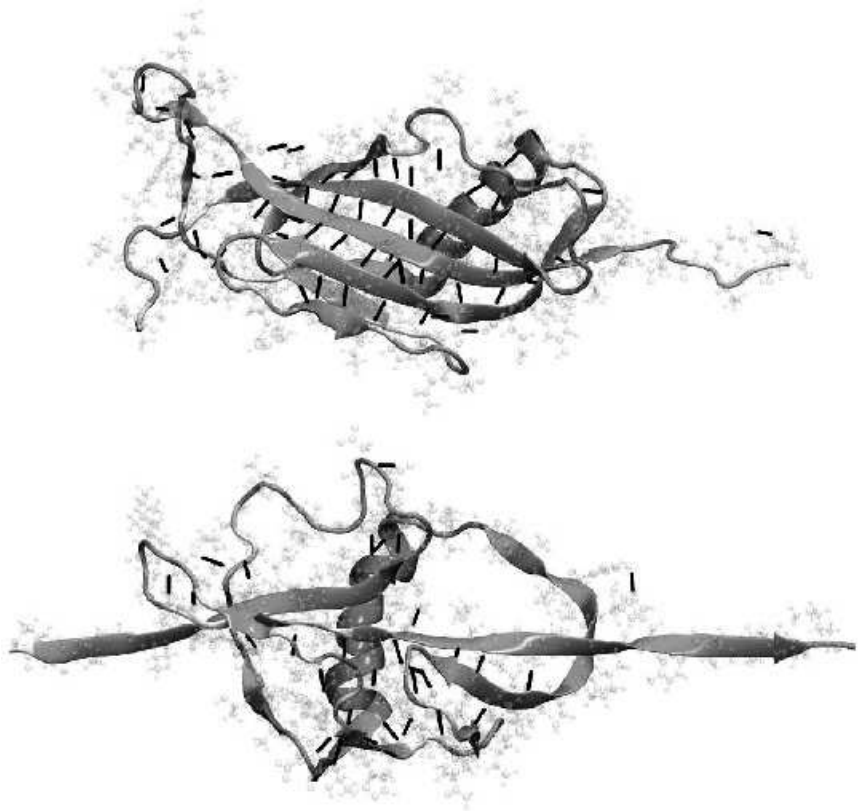}}
\vspace*{2cm}
\caption{{\bf Hydrogen bonds in 1c4p.} They are indicated by solid bars. They may link
the backbone or the side chains. The heavy atoms of the side chains are marked
in the faint way. The top panel corresponds to
$d$=20 {\AA} and the bottom panel to 80 {\AA}.
}
\label{hydro1c4p}
\end{figure}

The top left panel of Fig.\ref{wodor} and Fig.\ref{hydro1c4p}
illustrate the nearly monotonic
disappearance of the hydrogen bonds in 1c4p as a result of pulling.
At $F_{max}$, there are no obvious indicators of rupture in the total number
of the hydrogen bonds. However, such indicators arise when one looks into
the number of the hydrogen bonds in specific pairs of the strands (Fig.\ref{wodor}).
It is seen that the pairs $\beta_1$-$\beta_4$ and $\beta_1$-$\beta_2$ lose
their couplings at $F_{max}$, though some of them  reappear temporarily
later. On the other hand, the    
sheet $\beta_3$-$\beta_4$ 
loses fewer couplings
at $F_{max}$ suggesting
their stabilizing role. They are still seen as partly operational in the
lower panel of Fig.\ref{hydro1c4p} which shows the placement of the
hydrogen bonds at $d$=80 {\AA}. At this stage, the hydrogen bonds in the
15-residue long $\alpha$-helix remain nearly intact.

The mechanical behavior of the remaining proteins shown in Fig.\ref{shear}
is similar
to that found in 1c4p even though the corresponding structures are different and
the values of $F_{max}$ are smaller. Here, we discuss the diadenosine tetraphosphate
hydrolase 1f3y for which the maximum force (of 2553 -- 2624 pN)
arises not at the first but at the third force peak.
There are five helices in 1f3y (53-65, 85-95, 116-118, 139-145, and 148-164)
and seven  $\beta$-strands ($\beta 3$ is 36-41, $\beta 4$ is 44-46, and the
remaining ones are listed in the caption to Fig.\ref{wodor}.
The two terminal ones unfold at the beginning and generate a small force peak at
$d$ around 50 {\AA}. These events are followed by rupturing
the sheet formed by the antiparallel strands $\beta 2$ and $\beta 7$.
The second small peak (around $d$=160 {\AA}) comes from
unfolding the bonds related to residues in the segment between 113 and 130
which involves loops and the third short helix.
Panel (b) of Fig.\ref{wodor} demonstrates that the primary
shear mechanical clamp is formed by two parallel strands $\beta 1$ and $\beta 6$
with some contribution from the antiparallel $\beta 5$ and $\beta 6$.
The bonds between $\beta 2$ and $\beta 7$ are seen to disappear much earlier.

\subsection{The cystine slipknot  mechanical clamp}

\subsubsection{The $F-d$ patterns}
We can distinguish two characteristic groups of the $F-d$ curves: with one
force peak (1fzv, 1qty, 1cz8, 1wq9, 1flt, 1vpf, 1wq8) and with multiple
peaks. This feature originates primarily from the size of the slipknot loop.
In all of the single maxima cases, the proteins have a short
and stiff slipknot loop comprising of order six amino acids.
Larger slipknot loops (typically of order 30 amino acids) 
are more flexible and unfavorable steric clashes with the cytine ring get
split into several events in various ways and are protracted in time as
seen in Figs.\ref{three} and \ref{next}. Such long loops arise in
1bmp, 1lxi, 1m4u, 1rew, 2bhk, 2gh0, 2gyz, and 3bmp.
However, multiple peaks may sometimes also arise with short loops, as discussed
in the next section. This effect relates to sizes of the amino acids in
the loop.

We first consider two examples of proteins
for which we have observed only single force peaks: placenta
growth factor 1fzv and vascular endothelial growth factor 1vpf (both human).
In the case of 1fzv, the ring is formed from eight amino acids
Cys66--Thr67--Gly68--Cys69--Cys70--Cys113--Glu112--Cys111 that make a ring
because of the disulfide bonds Cys66-Cys111 and Cys70-Cys113.
The slip-loop  involves the segment
Leu75--His76--Cys77--Val78--Pro79 (see also Table 2).
Dragging is carried out by the disulfide bond Cys35-Cys77.
In the case of 1vpf, the ring is created by the disulfide
bonds Cys57-Cys102 and Cys61-Cys104 that link the backbone segments
Gly58--Gly59--Cys61 and Glu103 into a tight ring.
The slip-loop is formed by another segment of the backbone which
consists of amino acids Leu65--Glu67--Cys68--Val69--Pro70 (see also Table 2).
Dragging is carried out by the Cys36-Cys68 disulfide bond.

\begin{figure}[!ht]
\epsfxsize=4.0in
\centerline{\epsffile{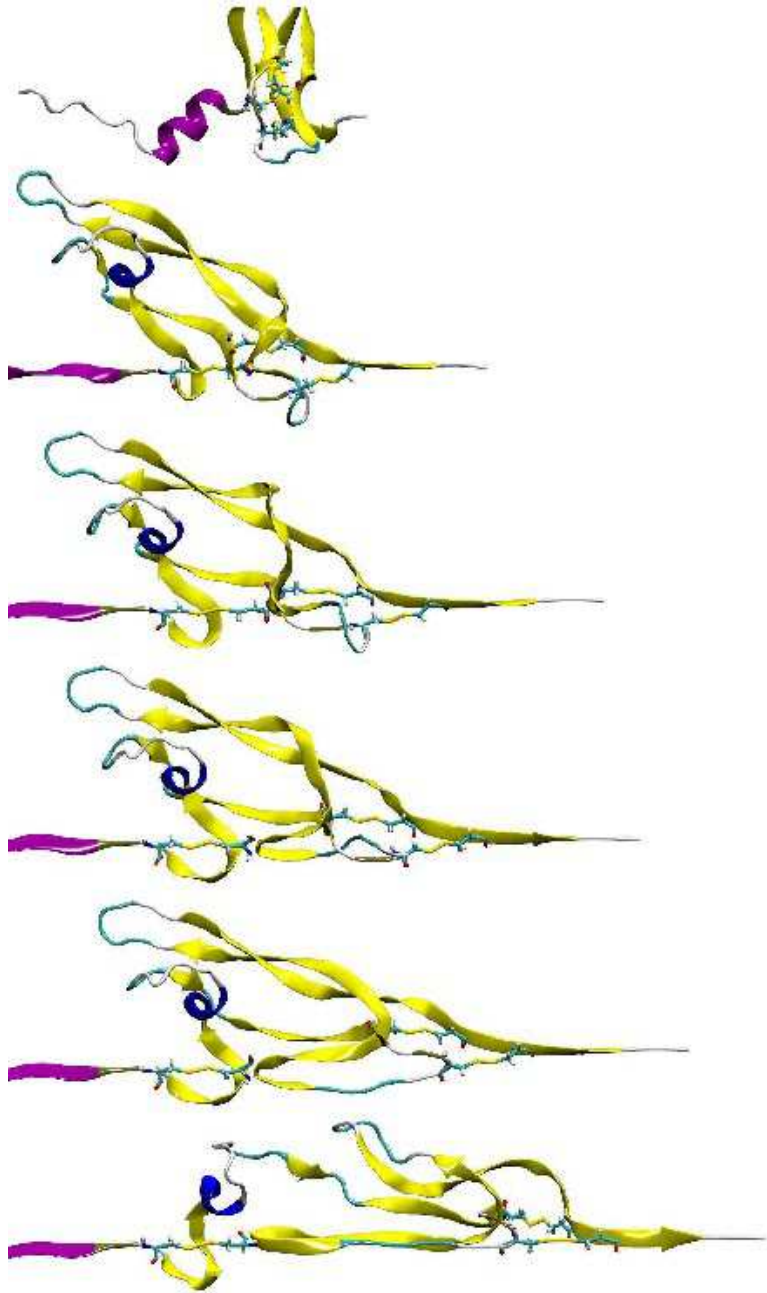}}
\vspace*{2cm}
\caption{{\bf Stages of rupturing of the cystine slipknot mechanical
clamp in protein 1fzv.}
The time frames correspond to $d$ of
0, 61, 113, 114, 114.5, and 116 {\AA} top to bottom respectively.
At the last stage shown here, the tension is decreasing even though
the slip-loop is not yet fully pulled through. The biggest jamming is at
$d$=114 {\AA}.
}
\label{stages}
\end{figure}

\begin{figure}[!ht]
\epsfxsize=5in
\centerline{\epsffile{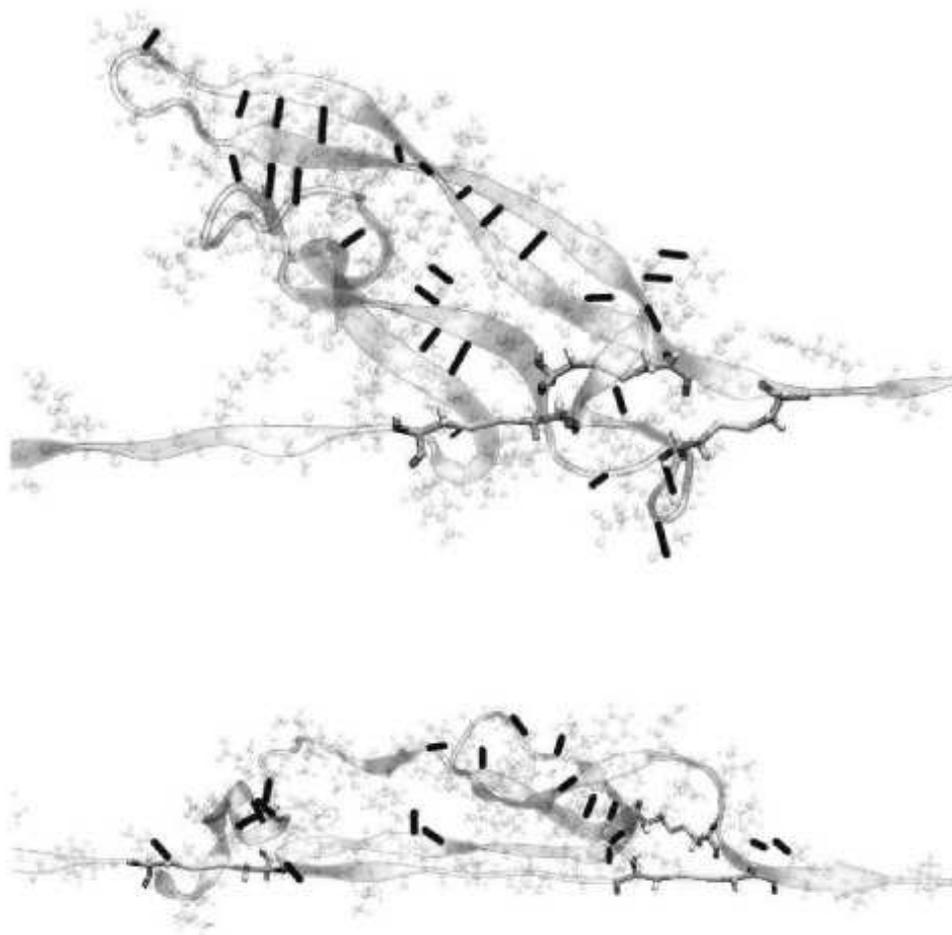}}
\caption{{\bf Hydrogen bonds in 1fzv for $d$=110 {\AA} (the top panel)
and 120 {\AA} (the bottom panel).}  Like in Fig.\ref{hydro1c4p}, they are
indicated as solid bars. The force peak arises at 114 {\AA} on this trajectory.
}
\label{hydro1fzv}
\end{figure}

The stages of unfolding of 1fzv are illustrated in Fig.\ref{stages} for the
trajectory with $F_{max}$ of 30 390 pN (the other one corresponds to 26 980 pN).
Dragging through the cystine ring is seen to take place abruptly -- it
is accomplished when $d$ varies by less than about 3 {\AA}. In the native state,
the slip-loop is stabilized by 12 hydrogen bonds between two long parallel
$\beta$-strands (amino acids 75--91 and 99--115).
Almost all of them get ruptured on
dragging the slip-loop through the ring as illustrated in Fig.\ref{hydro1fzv}.
We have demonstrated within the coarse grained model \cite{plos2009}
that replacing all of the slip-knot related contacts
(inside the slip-loop and between the slip-loop and the ring)
by repulsive contacts does not affect $F_{max}$ in a significant manner,
indicating that the CSK mechanical clamp operates mostly through
overcoming the sterical constraints.
Panel (c) in Fig.\ref{wodor} shows that the total number
of hydrogen bonds undergoes a sudden drop when the CSK clamp gets
ruptured because of a major transformation of the whole structure.
Panel (d) in this figure shows that the same observation
also applies to the multiple-peak CSK proteins like 1bmp. This has
not been the case in the shear mechanical clamps (panels (a) and (b) of
Fig.\ref{wodor}) because the unraveling at $F_{max}$ is
relatively more local and much more gentle.

\begin{figure}[!ht]
\epsfxsize=4in
\centerline{\epsffile{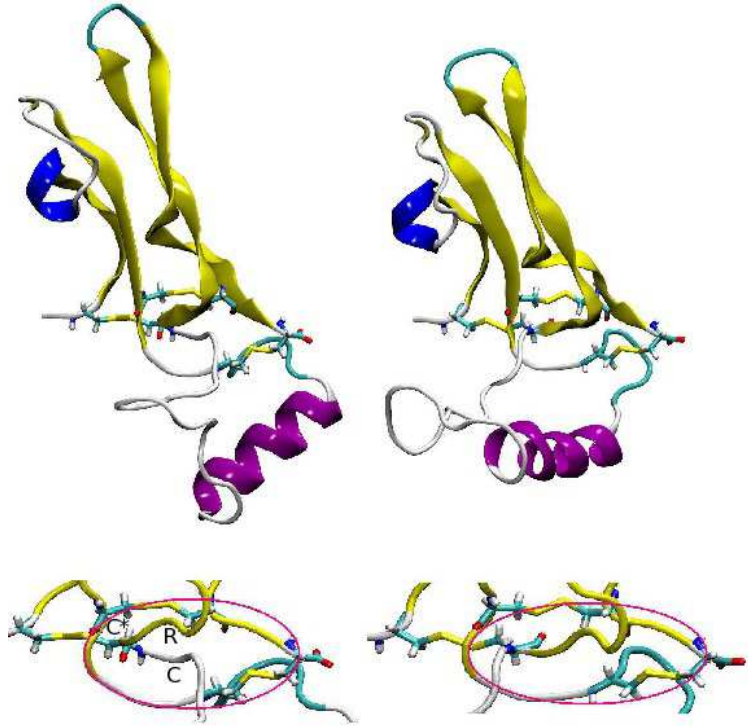}}
\vspace*{2cm}
\caption{{\bf The action of the cystine slipknot in 2gyz  on two
pathways.} The left panels correspond to the pathway with the
smaller value of $F_{max}$. The right panels -- to the pathway with
the larger $F_{max}$. All of the panels correspond to $d$=22 {\AA}.
In the lower panels, C$^*$ indicates the cysteine which is pulled and
C the neighboring cysteine. R denotes the neighboring arginine.
In the left panels C, is closer to the ring than R. In the right panels,
it is the other way around.
}
\label{slabas}
\end{figure}

\begin{figure}[!ht]
\epsfxsize=4in
\centerline{\epsffile{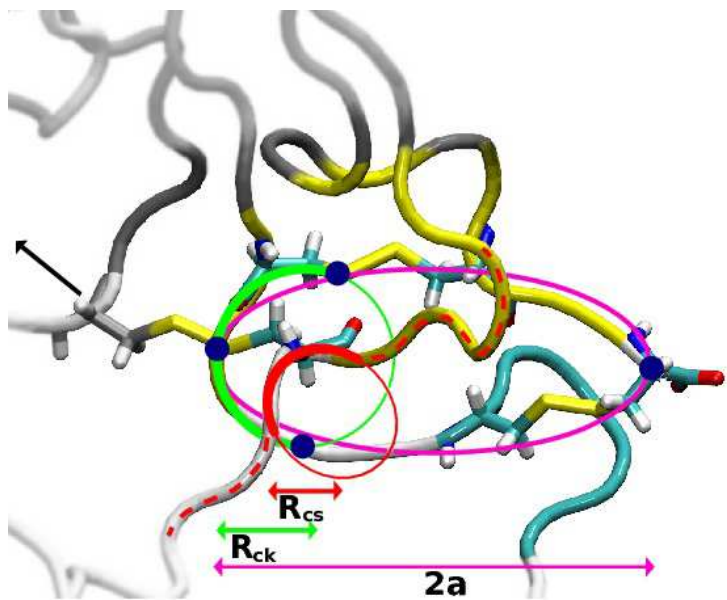}}
\vspace*{2cm}
\caption{{\bf Definition of the characteristic radii of curvature in
the CSK problem as illustrated for 2gyz in its native state.}
The ellipse in pink encompasses the cystine ring.
This loop is closed by two disulfide bonds Cys32-Cys98 and Cys36-Cys100.
The major semi-axis of the ellipse
is denoted by $a$ and is estimated by taking half of the distance
between the C$^{\alpha}$ atoms on the cystine ring which are furthest away
from each other (Cys70 and Cys98).
The segment near the "perihelion" is approximated by a circle in green.
Its curvature can be determined from a triangle formed by three
consecutive amino acids Ser33, Gly34 and Ser35.
The corresponding radius of curvature is
denoted by $R_{ck}$. Here, however, the circle shown is set on the
triangle based on the two consecutive C$^{\alpha}$ atoms in Ser33 and Gly34
with the third corner of the triangle placed at the S atom belonging to
the fourth amino acid Cys36. In the native
state, there is very little difference in the radii generated from these
two kinds of the triangles.
The circle in red, of radius $R_{cs}$, corresponds to the curvature obtained
by considering the three consecutive C$^{\alpha}$: Cys69, Cys70 and Arg71.
The middle C$^{\alpha}$
belongs to a cysteine.
This red circle is above the ellipse. On pulling, it gets squeezed and
dragged down by Cys70 through the ellipse.
}
\label{ellipse}
\end{figure}

We now consider the neurotrophic growth factor 2gyz. It is an example of a
CSK protein which has (at least) two unfolding pathways, each with several
force peaks (another such example is 1m4u).
On one pathway $F_{max}$ is nearly twice as large as on the
other. The cystine ring consists of eight amino acids:
Cys32--Ser33--Gly34--Ser35--Cys36--Cys100--Gly99--Cys98--Cys32.
It is indicated by an ellipse in Figs.\ref{slabas} and \ref{ellipse}.
Fig.\ref{slabas} shows two possible ways of crossing
the ring by the pulled cysteine whereas Fig.\ref{ellipse},
corresponding to the right-hand panels of Fig.\ref{slabas},
defines the geometrical parameters of the CSK that will be discussed
when  describing temporal changes in the structure of the
CSK mechanical clamp.
The slip-loop is long and is pulled by Cys70 which
forms a disulfide bond with Cys5 which is just near the N-terminus.
One branch of the slip-loop is 27-residue long (from Arg37 to Cys70)
and the other is 33-residue long (from Cys70 to Ala97).
The relevant segment of the slip-loop that has to squeeze through the
ring is then Gln67--Pro68--Cys69--Cys70--Arg71--Pro72--Thr73.
The steric hindrance arises when this segment of the slip-loop
passes near the Cys32--Ser33--Gly34 fragment of the ring.
Depending on the history of thermal fluctuations, there are only two
ways in which the slip-loop is facing the ring: either
the first neighbor Cys32 is closer to the ring or the other
first neighbor Arg97 is the one which is closer. The two ways
differ by a rotation of 180$^o$.
If the approach to the plane of the ring the were vertical, the two amino acids
would enter the ring simultaneously. Instead, the symmetry is
broken because the approach is at an acute angle since the pulling
direction becomes increasingly parallel to the longer axis of the
ellipse that represents the narrowing ring.
The first way,
shown in the left panel of Fig.\ref{slabas}, yields the smaller force of
15 nN since cysteine is the smaller sized amino acid of the two.
The second way, shown in the right panel, yields $F_{max}$ of 30 nN.
The small force case must necessarily come with the striated $F-d$
pattern since the bigger amino acid must follow immediately next
which involves further forced expansion of the ring.

The separation into large force and small force pathways, independent of
the number of the force peaks, should be observed for each of the
CSK proteins studied here. However, for short slip-loops one way is expected
to dominate, as determined by its sequential makeup,
but both possibilities are seen in stretching of 1vpf and 1wq9.
For long slip-loops, there is no particular preference
for selection of the initial orientation on approaching the ring.
It should be noted that the role of fluctuations depends on the
speed of pulling. At small pulling speeds,
conformations are expected to get equillibrated better which
should reduce choices between various trajectories.

\subsubsection{Residue-based description of jamming}

\begin{figure}[!ht]
\epsfxsize=4in
\centerline{\epsffile{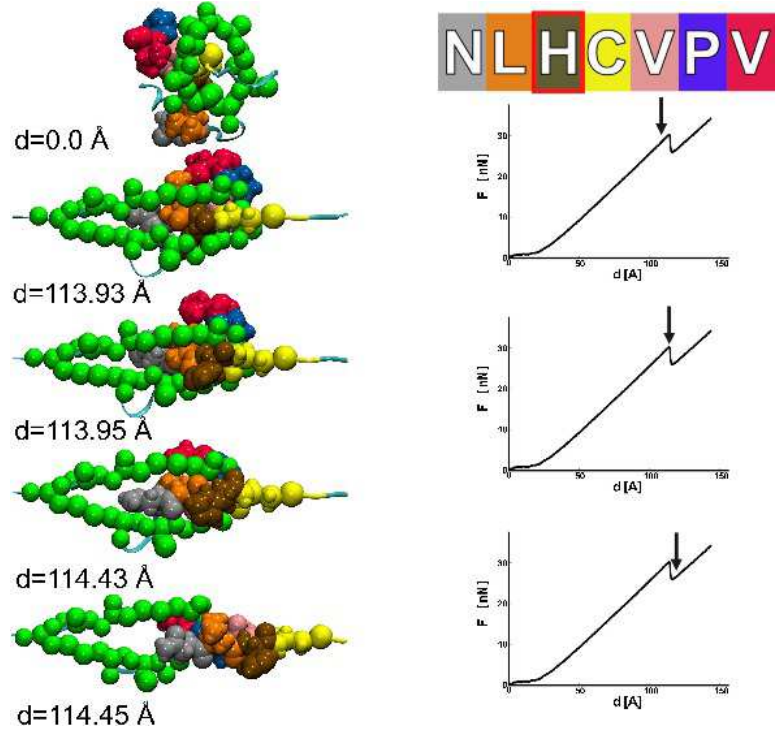}}
\vspace*{2cm}
\caption{{\bf The action of the cystine slipknot in 1fzv
in the strong force pathway.}
The beads represent van der Waals spheres associated with the
heavy amino acids. The different colors correspond to the
amino acids indicated at the top.
The amino acid that enters
the cystine ring right after the cysteine is marked
by the rectangular frame  in the pictorial representation of the sequence.
The snapshots correspond to the
instances marked by the arrows on the $F-d$ curves on the right
show situation just before sudden drop in tension.
}
\label{cleat}
\end{figure}

\begin{figure}[!ht]
\epsfxsize=5in
\centerline{\epsffile{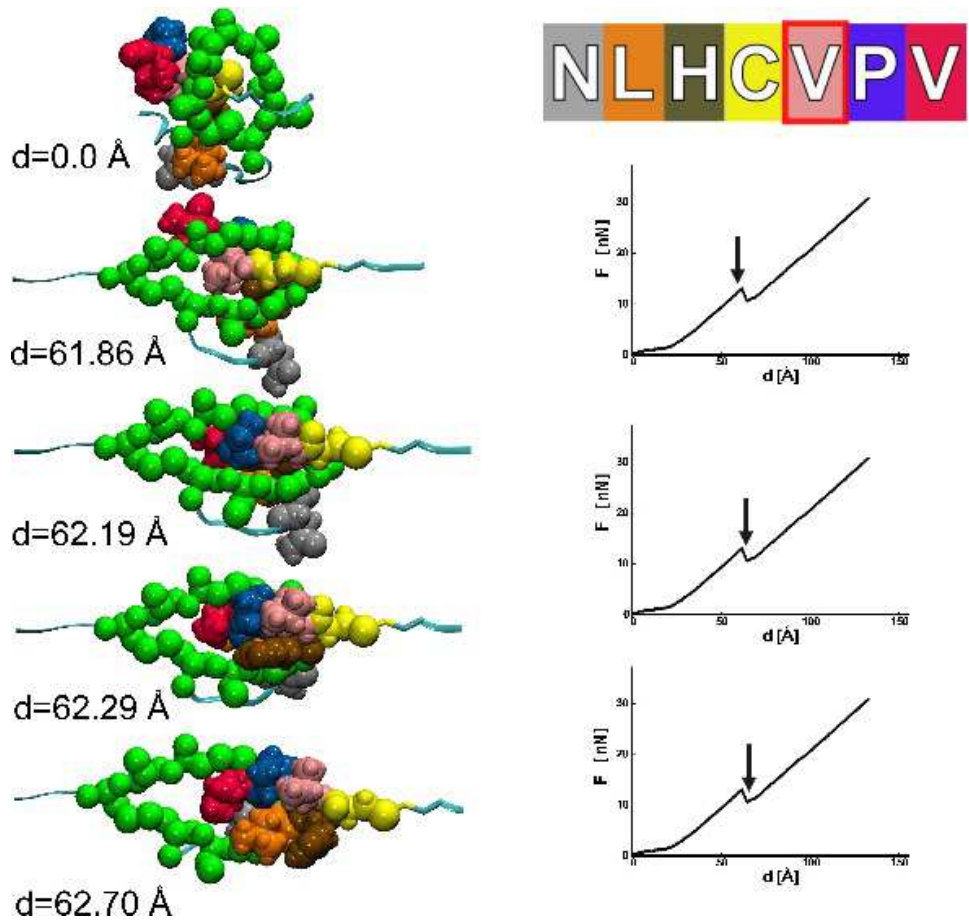}}
\vspace*{2cm}
\caption{ {\bf The action of the cystine slipknot in 1fzv
in the "weak" force pathway.} The figure is similar to Fig.\ref{cleat}
but it corresponds to Val78 entering the first. The trajectory chosen
for illustration is the one with the smallest $F_{max}$ in Fig.\ref{next}g.
}
\label{cleat1}
\end{figure}

We first consider proteins with short slipknot loops such as 1fzv and 1vpf.
Figs.\ref{cleat} and \ref{cleat1} illustrate the jamming process through the evolution
of the clinched fragment of 1fzv in the immediate vicinity of the force
peak. The relaxed, native structure of the fragment is shown at the top
for comparison. The fragment comprises the full cystine ring (in green)
and five amino acids of the slip-loop near its dragging head -- 
Cys75 (in yellow).  The three-residue long fragments
of the slip-loop on the left and on the right of the pulling
cysteine are listed in Table 2. The table makes use of the
sequential alignments as in refs. \cite{kretchmer,Craik2,pallaghy}.
Note that for each of the
entries listed in the table, the second sequential neighbor of the
pulling cysteine, when counting away from the N-terminus, is always a proline
which initiates a $\beta$-sheet.
In all but three cases, the proline is followed by threonine. Between the pulling
cysteine and the proline, the residues can be either small or large.

The cysteine is a relatively small amino
acid so it slides into the cystine ring fairly easily.
This step, however, bends the slip-loop and raises tension.
On further pulling, the neighboring residues (His76 and Val78 shown
in brown and pink in the figures respectively) have to
penetrate the ring either His76 first or Val78 first with
the other of the two following immediately afterwards.
The pathway with His75 entering first is dominant because the
corresponding conformation is more native-like. It is shown in
Fig.\ref{cleat} whereas the other pathway -- in Fig.\ref{cleat1}.
The penetration is resisted by the ring and the system is jammed.
This stage determines the value of $F_{max}$. The dominant pathway
yields a larger $F_{max}$ since histidine is bigger in size than valine
because of its aromatic ring. In this pathway, when His76 stays jammed,
its neighbors Leu75 and  Asn74  together with Val78 (which is flanking Cys77
from the other side) form a compact complex. This complex has
an attached "hook", Pro79 (in blue), which is pivoting on the ring.
Immediately after His76
passes through the cystine ring, this complex passes the ring as well. This
event determines the largest force and is very quick. The whole process
generates a single maximum. Once the bulky Leu75, Asn74 and Val78 complex
crosses the loosened ring, the remaining slipknot residues slide through
easily in a swinging motion.
When the full length of the slip-loop is exhausted and stretching affects
only covalent bonds from the protein backbone, the tension grows again.
The pathway in which Val78 is the head amino acid which penetrates the
cystine ring (Fig.\ref{cleat1}) is similar with the role of the
neighboring amino acids interchanged. For instance, it is Leu75 which
forms a hugging hook on the ring now.

The single peak mechanism involving a passage of a bulky plug is also
observed in 1qty, 1wq8, 1flt, 1fzv, 1vpf, and 1wq9. In each of
these proteins, the other second neighbor of the pulling cysteine
is bulkier than proline and, in the dominant pathway, it is the larger
second residue which penetrates the first and paves the way for the
smaller companion. It should be pointed out that
despite the multitude of the possible values of $F_{max}$, as seen,
e.g. in Fig.\ref{next}g, there are only two pathways of the
ring penetration. The differences in the values arise merely from
minute differences in the angle of approach that get emphasized
by the very magnitude of the stresses involved.

Multiple peaks arise when the slipknot loop
is long and when the amino acids
enter in order in which their geometrical sizes keep increasing
because the subsequent
residues keep forcing the cystine ring ever more open.
The resulting sawtooth-like $F-d$ pattern is encountered, for instance,
in 1bmp, 2gyz, 2bhk, 2gh0, 1rew, 3bmp. Again, there are basically only two pathways,
albeit multipeaked.
It is interesting to observe that all of these multiple-peak
proteins contain another cysteine. It should be noted that this
neighboring cysteine may be involved in forming a dimer state with
a similar companion. In the dimeric state, the slip-loop motion
through the ring would be prohibited.

We have found that the passage of the slipknot through the
cystine ring is generally associated with a rapid rise in the bond and
bond-angle contributions to the total energy. All remaining contributions
to the total energy, such as electrostatic, van der Waals and those
associated with the dihedral and improper angles remain
fairly unaffected by crossing of the force peaks.

\subsubsection{Curvatures in the backbone}
Finally, we discuss the geometry of the CSK mechanical clamp
in terms of geometry as characterized by the effective curvatures.
Previously \cite{plos2009}, we have proposed that the slip-loop
can be driven through the cystine ring provided
\begin{equation}
R_{cs}+t_s \;<\; R_{ck}-t_k\;\;\;,
\label{condition}
\end{equation}
where $R_{cs}$ and $R_{ck}$ are the radii of curvature of the slip-loop
and the ring respectively. We have estimated these radii to be of order
7 and 3 {\AA} correspondingly.
Symbol $t_s$ denotes the effective thickness
of the slip-loop and $t_k$ that of the ring. Both have been assessed
to be around 2.5 {\AA}.

\begin{figure}[!ht]
\epsfxsize=3in
\centerline{\epsffile{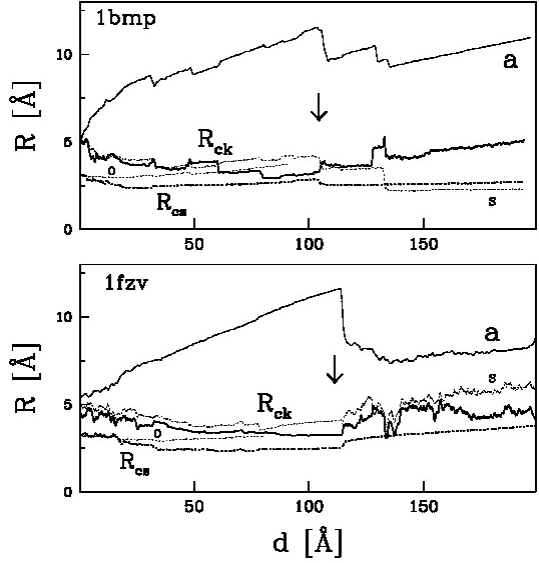}}
\vspace*{2cm}
\caption{{\bf The evolution of the characteristic radii of curvature and
of the parameter $a$ in 1bmp (top) and 1fzv (bottom).} The arrows
indicate the location of the highest force peak on the trajectory analyzed.
The curvatures determined based on the three consecutive C$^{\alpha}$ atoms
are marked by the thicker lines. The solid line is for the
cystine ring and the dotted line -- for the slip-loop.
The thin line marked by the symbol S
is obtained when the third C$^{\alpha}$,
belonging to the cysteine in the cystine ring,
is replaced by the sulphur atom on the fourth amino acid. This atom
forms one of the disulfide bonds defining the cystine ring.
The thin line marked by the symbol O is obtained when the
forward C$^{\alpha}$ in the slip-loop
(i.e. in the amino acids that is first pulled through the ring)
is replaced by the associated backbone oxygen atom.
}
\label{curvature}
\end{figure}

Here, we reexamine this condition by determining the local radii
of curvature in a way used in the context of the tube picture
of proteins \cite{tube,tube1}, i.e. by finding a circle which
goes through the corners of a triangle set by three consecutive
C$^{\alpha}$ atoms. Fig.\ref{curvature} shows the thus determined
$R_{cs}$ and $R_{ck}$ for 1bmp and 1fzv, together with the semimajor
axis, $a$, of the ellipse assigned to the cystine ring. In the
native state, $a$ and $R_{ck}$ are both close to 5 {\AA} for both proteins.
$R_{cs}$ is found to be about 3 {\AA}. On pulling, all of these
parameters evolve. Until reaching $F_{max}$, $a$ generally grows
whereas $R_{ck}$ and $R_{cs}$ go down. At $F_{max}$, the radii
jump upward, indicating an escape from the constriction, and $a$
jumps downward. At each stage, the condition given by eq. \ref{condition}
is satisfied if the tube thickness is neglected.

In order to estimate the effect of the thickness
in a description which does not invoke an explicit tube, we replace one of
the corners of the triangle by another atom. For a new effective
$R_{cs}$ we take the backbone oxygen atom associated with the
pulled amino acid (the line denoted by O). For a new effective
$R_{ck}$ we take the sulphur atom (the line denoted by S) residing
on the next coming amino acid (see also Fig.\ref{ellipse}).
These displaced-atom lines generally follow the behavior of the
C$^{\alpha}$ based curvatures, at least until reaching $F_{max}$.
However, they illustrate the magnitude of the effects associated
with the tube thickness. The shifts are seen to be substantial and
indicate that the condition to go through is borderline or even
prohibitive if viewed from the tube-like perspective.
We conclude that the atomic-level features and structure
are important for the CSK mechanism to work when $R_{cs}$ and $R_{ck}$
are comparable in size.


\vspace*{0.5cm}
\subsection{Concluding remarks}

Our all-atom simulations confirm existence of huge mechanostability
of the CSK proteins that has been predicted by using the coarse grained model.
It should be pointed out that there are differences between the way
the slip-knot passes through the cystine ring as described by the two models.
The all-atom ring is much tighter because of the  bigger excluded volume  generated by
the atoms in the side groups. Thus the passage process involves significant
distortion of the ring so that it becomes a narrow ellipse. As a result,
only one amino acid can squeeze through at a time. In contrast, the C$^{\alpha}$-built
ring in the coarse-grained model remains fairly circular allowing for the
simultaneous passage of up to two amino acids that are the first neighbors
of the pulling cysteine. The variations in the
magnitude of $F_{max}$ between the trajectories in the coarse grained model,
also even up to a factor of 2, arise  now not that much from the selection of the amino
acid that goes through first but from the possibility of dragging of one
or two neighboring amino acids initially.

Our all-atom simulations validate the results obtained through coarse grained
modeling in a qualitative way as they support possible existence of the CSK mechanical
clamp. The ultimate test, however, should come from experimental pulling studies
and it is the experiments that should establish the true values of $F_{max}$.
Our theoretical estimates of $F_{max}$ are not sufficiently precise and should
be considered just as indicative of large forces that are expected to be found.
The difficulty that such experiments might encounter is the
resolvability of the CSK-related force peaks
from the background generated by stretching of the peptide bonds.

This stability of the CK proteins appears to be related to the
high mechanical stability found in this paper. Specifically,
we have shown that the buildup in the tension results from jamming.
The sequential neighbors of the pulling cysteine are of a size
that prevents them from penetrating the cystine ring without
forcing. Thus we expect that thermal fluctuations would
not be able to form a slipknot and change the conformation.
Similar stability features  should also be operational in the CCK
and ICK proteins. Thus even though mechanical stability, in general, need not
be related to stability against thermal fluctuations (see, e.g. \cite{thermunfold})
there is a relationship in the case of the CK proteins.
In particular, it has been shown \cite{Welfle} that a removal of the disulfide
bonds in the CK-containing VEGF protein through mutations lowers the
melting temperature significantly.

In a recent survey of multidomain proteins \cite{multi} still new mechanical clamps
have been identified by using the coarse grained model.
Among them are two kinds of tensile clamps -- one simple,
involving contacts between two domains, and another in which a part of the
clamp is immobilized by a non-cystine knot-loop. It would be interesting to
investigate atomic aspects of the workings of such clamps by using
all-atom simulations as done in this paper.


\section{Materials and Methods}

All simulations have been performed using NAMD 2.6 code \cite{Phillips}
with the all-atom CHARMM27 force field \cite{MacKerell}. All of the simulations have been
carried out using the same computational protocol.
The initial structure is downloaded from the PDB database \cite{PDB}
and then placed within a box with rigid water molecules. The layer of the water
molecules is at least 8 {\AA} thick. Na$^+$ and Cl$^-$ ions are introduced into
the system at concentration of 0.5 mol/L (between 10 and 20 ions in the box).
In order to neutralize the system, additional several ions of one sign are
added using a script embedded in the VMD code \cite{Humphrey}.
The molecules of water are then moved for 100 steps per atom
in order to minimize the energy and then equilibrated
at the temperature of 300 K during 100 ps of the time evolution.
After this stage, the protein atoms are allowed to
move and the whole system undergoes 1000 steps of the energy
minimization followed by heating
from 0 K to 300 K during 50 ps in a stepwise fashion.
The last step before stretching consists of 1 ns equilibration
at 300 K by using the Langevin dynamics.  The time step is set at 1 fs.

In analogy to the coarse-grained approach \cite{Robbins},
the SMD simulations \cite{Isralewitz,Isralewitz1,P2_t,Peplowski,Nowak}
are implemented by placing  the N-terminal C$^{\alpha}$ atom at
a fixed location  and by attaching an elastic
spring to the C$^{\alpha}$ atom at the C-terminal amino acid.
The elastic constant is equal to 4 kcal/mol/{\AA}$^2$ (277.9 pN/{\AA}).
The other end of the spring
is made to move with the pulling velocity $v_p$ of
0.04 {\AA}/ps (4 m/s). The pulling direction is set to go
through the line connecting the first and last C$^{\alpha}$ atoms.
In all cases, at least two 10 ns SMD simulations that evolve
from the same starting configuration are carried out. This is the configuration
obtained at the end of the initial Langevin dynamics. Various trajectories
arise due to differences in the time dependence of the Langevin noise applied.
Graphical analysis are performed with the VMD 1.8.7 software \cite{Humphrey}.
Forces and local geometries are analyzed by using homemade scripts.

In order to determine whether two heavy atoms are connected by
a hydrogen bond we check whether the distance between them
does not exceed 3.3 {\AA} and the angle donor (like O) -- H -- acceptor (like N)
does not exceed 25$^o$. These bonds rupture often and then reform.

The forces and the numbers of the hydrogen bonds are time
averaged over the pulling distances of 0.5 {\AA}. In the plots, these data
undergo further smoothing out.

\section{Acknowledgments}
Very useful discussions with J. I. Su{\l}kowska at the beginning
of the project are appreciated.
This work has been supported by the grant N N202 0852 33 from the Ministry
of Science and Higher Education in Poland, the EC FUNMOL project under
FP7-NMP-2007-SMALL-1 and through the Innovative Economy grant (POIG.01.01.02-00-008/08).


\clearpage
\section*{Figures}

\vspace*{1cm}

\vspace*{1cm}

\newpage
\clearpage
\section*{Tables}

TABLE 1: The predicted list of the strongest proteins.

\vspace*{0.5cm}
\begin{centering}
\begin{small}
\begin{xtabular}{|l|l|l|l|l|l|l|l|} \hline
n & PDBid & N & F$_{max}$ [pN] & $L_{max}$ [{\AA}]& $d_{max}$ [{\AA}] & CATH & SCOP \\ \hline
 1 &\bf{ 1bmp }& 104 &\bf{~1120 }&  23.2 & 176.0 & 2.10.90.10 & g.17.1.2 \\
 2 &\bf{ 1qty }& 95 &\bf{~ 980 }&  72.1 & 108.3 & 2.10.90.10 & b.1.1.4 \\
 3 &\bf{ 2bhk }& 119 &\bf{~ 800 }&  26.5 & 129.3 &   &   \\
 4 &\bf{ 1lxi }& 104 &\bf{~ 800 }&  22.5 & 126.4 &   & g.17.1.2 \\
 5 &\bf{ 1cz8 }& 107 &\bf{~ 700 }&  76.5 & 149.4 & 2.10.90.10 & b.1.1.1 \\
 6 &\bf{ 2gh0 }& 219 &\bf{~ 640 }&  25.9 & 104.6 &   &   \\
 7 &\bf{ 1wq9 }& 100 &\bf{~ 610 }&  72.0 & 131.8 & 2.10.90.10 & g.17.1.1 \\
 8 &\bf{ 1flt }& 107 &\bf{~ 610 }&  75.6 & 128.6 & 2.10.90.10 & b.1.1.4 \\
 9 &\bf{ 1fzv }& 117 &\bf{~ 590 }&  90.4 & 130.3 & 2.10.90.10 & g.17.1.1 \\
 10 &\bf{ 2gyz }& 100 &\bf{~ 590 }&  14.4 & 93.3 \&110 &   &   \\
 11 &\bf{ 1rew }& 103 &\bf{~ 580 }&  21.7 & 92.7 & 2.10.90.10 & g.7.1.3 \\
 12 &\bf{ 1m4u }& 139 &\bf{~ 580 }&  52.1 & 114.5 & 2.10.90.10 & g.17.1.2 \\
 13 &\bf{ 1vpf }& 94 &\bf{~ 580 }&  68.1 & 121.6 & 2.10.90.10 & g.17.1.1 \\ 
 14 &\bf{ 1c4p }& 137 &\bf{~ 560 }& 106.0 & 138.8 & 3.10.20.180 & d.15.5.1 \\
 15 &\bf{ 1qqr }& 138 &\bf{~ 550 }& 110.3 & 134.8 & 3.10.20.180 & d.15.5.1 \\
 16 &\bf{ 3bmp }& 114 &\bf{~ 550 }&  33.0 & 96.1 & 2.10.90.10 & g.17.1.2 \\
 17 &\bf{ 1j8s }& 193 &\bf{~ 540 }&  77.9 & 99.3 & 2.60.40.1370 & b.2.3.3 \\
 18 &\bf{ 1wq8 }& 96 &\bf{~ 540 }&  82.6 & 109.8 & 2.10.90.10 & g.17.1.1 \\
 19 &\bf{ 1j8r }& 193 &\bf{~ 530 }&  77.7 & 97.0 & 2.60.40.1370 & b.2.3.3 \\
 20 &\bf{ 1f3y }& 165 &\bf{~ 530 }& 284.7 & 336.0 & 3.90.79.10 & d.113.1.1 \\
3580 &\bf{ 1tit }&  89 &\bf{~ 230 }&  55.3 & 43.3 & 2.60.40.10 & b.1.1.4 \\
\end{xtabular}
\end{small}
\end{centering}

\vspace*{1cm}
Table 1.  $F_{max}$ is obtained within the structure-based coarse grained model
in ref. \cite{plos2009}. The model is defined in terms of the energy parameter
$\epsilon$ that determines the depth of the potential well in the native
contacts. The conversion to pN is by taking the average relationship
$\epsilon /${\AA} $\sim 110 $ pN.
The pulling velocity velocity is $\sim$0.005 {\AA}/ns.
The first column indicates the ranking of a model protein, the second --
the PDB code, and the third -- the number of the amino acids that are present
in the structure used.
$L_{max}$ denotes the end-to-end distance at which
the maximum force arises and $d_{max}$ the corresponding tip displacement.
The last two columns give the leading CATH and SCOP codes.

\clearpage

TABLE 2: Sequences of the pulled pieces of the slip-loops.
\\
\vspace*{0.5cm}
\hspace*{2cm}
\large
 \begin{minipage}[r]{0.5\linewidth} 

 \begin{xtabular}{|p{5.9cm}|p{1.7cm}|p{5cm}|} \hline
 Protein                                 & PDBid & Slip-loop sequence\\ \hline
 BMP-7                                   & 1bmp  &{\tt {\bf \colorbox{white}{\color{black}K}\colorbox{HPgray}{\color{white}P}\colorbox{white}{\color{black}C}\colorbox{CYSgray}{\color{white}C}\colorbox{white}{\color{black}A}\colorbox{HPgray}{\color{white}P}\colorbox{white}{\color{black}T} }}\\
 BMP-7                                   & 1m4u  &{\tt {\bf \colorbox{white}{\color{black}K}\colorbox{HPgray}{\color{white}P}\colorbox{white}{\color{black}C}\colorbox{CYSgray}{\color{white}C}\colorbox{white}{\color{black}A}\colorbox{HPgray}{\color{white}P}\colorbox{white}{\color{black}T}  }}\\
 BMP-7                                   & 1lxi  &{\tt {\bf \colorbox{white}{\color{black}K}\colorbox{HPgray}{\color{white}P}\colorbox{white}{\color{black}C}\colorbox{CYSgray}{\color{white}C}\colorbox{white}{\color{black}A}\colorbox{HPgray}{\color{white}P}\colorbox{white}{\color{black}T}  }}\\
 BMP-2                                   & 3bmp  &{\tt {\bf \colorbox{white}{\color{black}K}\scalebox{0.95}[1]{\colorbox{HPgray}{\color{white}A}}\colorbox{white}{\color{black}C}\colorbox{CYSgray}{\color{white}C}\colorbox{white}{\color{black}V}\colorbox{HPgray}{\color{white}P}\colorbox{white}{\color{black}T}  }}\\
 BMP-2                                   & 1rew  &{\tt {\bf \colorbox{white}{\color{black}K}\scalebox{0.95}[1]{\colorbox{HPgray}{\color{white}A}}\colorbox{white}{\color{black}C}\colorbox{CYSgray}{\color{white}C}\colorbox{white}{\color{black}V}\colorbox{HPgray}{\color{white}P}\colorbox{white}{\color{black}T}  }}  \\
 VEGF                                    & 1qty  &{\tt {\bf \colorbox{white}{\color{black}G}\scalebox{1.09}[1]{\colorbox{HPgray}{\color{white}L}}\hspace*{-0.07cm}\scalebox{1.06}[1]{\colorbox{white}{\color{black}E}}\colorbox{CYSgray}{\color{white}C}\colorbox{white}{\color{black}V}\colorbox{HPgray}{\color{white}P}\colorbox{white}{\color{black}T}  }} \\
 VEGF                                    & 1cz8  &{\tt {\bf \colorbox{white}{\color{black}G}\scalebox{1.09}[1]{\colorbox{HPgray}{\color{white}L}}\hspace*{-0.07cm}\scalebox{1.06}[1]{\colorbox{white}{\color{black}E}}\colorbox{CYSgray}{\color{white}C}\colorbox{white}{\color{black}V}\colorbox{HPgray}{\color{white}P}\colorbox{white}{\color{black}T}  }} \\
 VEGF                                    & 1flt  &{\tt {\bf \colorbox{white}{\color{black}G}\scalebox{1.09}[1]{\colorbox{HPgray}{\color{white}L}}\hspace*{-0.07cm}\scalebox{1.06}[1]{\colorbox{white}{\color{black}E}}\colorbox{CYSgray}{\color{white}C}\colorbox{white}{\color{black}V}\colorbox{HPgray}{\color{white}P}\colorbox{white}{\color{black}T}  }}  \\
 VEGF                                    & 1vpf  &{\tt {\bf \colorbox{white}{\color{black}G}\scalebox{1.09}[1]{\colorbox{HPgray}{\color{white}L}}\hspace*{-0.07cm}\scalebox{1.06}[1]{\colorbox{white}{\color{black}E}}\colorbox{CYSgray}{\color{white}C}\colorbox{white}{\color{black}V}\colorbox{HPgray}{\color{white}P}\colorbox{white}{\color{black}T}  }}  \\
 VEGF                                    & 1wq9  &{\tt {\bf \scalebox{1.22}[1]{\colorbox{white}{\color{black}S}}\hspace*{-0.05cm}\scalebox{0.82}[1]{\colorbox{HPgray}{\color{white}M}}\scalebox{0.96}[1]{\colorbox{white}{\color{black}K}}\colorbox{CYSgray}{\color{white}C}\scalebox{1.05}[1]{\colorbox{white}{\color{black}T}}\colorbox{HPgray}{\color{white}P}\colorbox{white}{\color{black}V}  }}   \\
 VEGF toxin                              & 1wq8  &{\tt {\bf \scalebox{1.22}[1]{\colorbox{white}{\color{black}S}}\hspace*{-0.05cm}\scalebox{0.82}[1]{\colorbox{HPgray}{\color{white}L}}\scalebox{0.96}[1]{\colorbox{white}{\color{black}K}}\colorbox{CYSgray}{\color{white}C}\scalebox{1.05}[1]{\colorbox{white}{\color{black}T}}\colorbox{HPgray}{\color{white}P}\colorbox{white}{\color{black}V}  }}  \\
 Human GDF                               & 2bhk  &{\tt {\bf \scalebox{1.08}[1]{\colorbox{white}{\color{black}P}}\scalebox{1.00}[1]{\colorbox{HPgray}{\color{white}T}}\colorbox{white}{\color{black}C}\colorbox{CYSgray}{\color{white}C}\colorbox{white}{\color{black}V}\colorbox{HPgray}{\color{white}P}\colorbox{white}{\color{black}T}  }}\\
 GDNF family receptor $\alpha$-3         & 2gh0  &{\tt {\bf \scalebox{1.08}[1]{\colorbox{white}{\color{black}Q}}\scalebox{1.00}[1]{\colorbox{HPgray}{\color{white}P}}\colorbox{white}{\color{black}C}\colorbox{CYSgray}{\color{white}C}\colorbox{white}{\color{black}R}\colorbox{HPgray}{\color{white}P}\colorbox{white}{\color{black}T}  }}\\
 PlGF                                    & 1fzv  &{\tt {\bf \colorbox{white}{\color{black}N}\hspace*{-0.05cm}\scalebox{1.09}[1]{\colorbox{HPgray}{\color{white}L}}\scalebox{0.93}[1]{\colorbox{white}{\color{black}H}}\colorbox{CYSgray}{\color{white}C}\colorbox{white}{\color{black}V}\colorbox{HPgray}{\color{white}P}\colorbox{white}{\color{black}V}  }}   \\
 ARTN isoform 3                          & 2gyz  &{\tt {\bf \scalebox{1.02}[1]{\colorbox{white}{\color{black}Q}}\colorbox{HPgray}{\color{white}P}\colorbox{white}{\color{black}C}\colorbox{CYSgray}{\color{white}C}\colorbox{white}{\color{black}R}\colorbox{HPgray}{\color{white}P}\colorbox{white}{\color{black}T}  }} \\
 \hline
 \end{xtabular}
 \end{minipage}

\vspace*{1cm}
Table 2. Abbreviations used in the table:
BMP - Bone morphogenetic protein,
VEGF - Vascular endothelial growth factor,
GDF - growth and differentiation factor,
GDNF - Glial cell line-derived neurotrophic factor,
PlGF - Placenta growth factor,
ARTN - Neurotrophic factor artemin.

\end{document}